\documentclass[aps,prb,reprint,showpacs,superscriptaddress,longbibliography]{revtex4-2}
\usepackage{mathtools,amssymb,graphicx,units,bm}
\usepackage{amsmath}
\usepackage[plainpages=false,pdfpagelabels,colorlinks=true,linkcolor=red,urlcolor=blue,citecolor=blue,pdftitle={Title},pdfauthor={},pdfdisplaydoctitle=true,pdfduplex=DuplexFlipLongEdge]{hyperref}
\usepackage{soul} 
\usepackage[dvipsnames,usenames]{color}
\usepackage[normalem]{ulem}
\usepackage{graphicx}
\graphicspath{{figures/}}
\usepackage{color}
\usepackage{bbold} 
\usepackage{float}
\emergencystretch=\maxdimen
\usepackage{graphicx}
\usepackage{dcolumn}
\usepackage{bm}
\usepackage{physics}

\def\ave#1{\langle #1\rangle}

\begin{document}

\title{Topological marker approach to an interacting Su-Schrieffer-Heeger model}

\author{Pedro B. Melo}
 \email{pedrobmelo@aluno.puc-rio.br}
\affiliation{Departamento de F\'\i sica, Pontif\'\i cia Universidade Cat\'{o}lica do Rio de Janeiro, 22452-970 Rio de Janeiro RJ, Brazil}
\affiliation{Instituto de F\'\i sica, Universidade Federal do Rio de Janeiro
Cx.P. 68.528, 21941-972 Rio de Janeiro RJ, Brazil}
\author{Sebastião A. S. Júnior}%
\affiliation{Instituto de F\'\i sica, Universidade Federal do Rio de Janeiro
Cx.P. 68.528, 21941-972 Rio de Janeiro RJ, Brazil}
\author{Wei Chen}
\affiliation{Departamento de F\'\i sica, Pontif\'\i cia Universidade Cat\'{o}lica do Rio de Janeiro, 22452-970 Rio de Janeiro RJ, Brazil}
\author{Rubem Mondaini}
\affiliation{Beijing Computational Science Research Center, Beijing 100193, China}
\author{Thereza Paiva}
\affiliation{Instituto de F\'\i sica, Universidade Federal do Rio de Janeiro
Cx.P. 68.528, 21941-972 Rio de Janeiro RJ, Brazil}

\begin{abstract}
The topological properties of the Su-Schrieffer-Heeger (SSH) model in the presence of nearest-neighbor interaction are investigated by means of a topological
marker, generalized from a noninteracting one by utilizing the single-particle Green's function of the many-body ground state. We find that despite the marker not being perfectly quantized in the presence of interactions, it always remains finite in the topologically nontrivial phase while converging to zero in the trivial phase when approaching the thermodynamic limit, and hence correctly judges the topological phases in the presence of interactions. The marker also correctly captures the interaction-driven, second-order phase transitions between a topological phase and a Landau-ordered phase, which is a charge density wave order in our model with a local order parameter, as confirmed by the calculation of entanglement entropy and the many-body Zak phase. Our work thus points to the possibility of generalizing topological markers to interacting systems through Green's function, which may be feasible for topological insulators in any dimension and symmetry class.
\end{abstract}

\maketitle

\section{Introduction}

In the research of topological insulators (TIs) and topological superconductors (TSCs), a particularly important issue is whether our understanding of the topological order in noninteracting systems still remain true in the presence of various correlations since they generally occur in real materials~\cite{Rachel18}. When interactions are absent, our current knowledge of the topological order within the context of Dirac models is fairly complete~\cite{Schnyder08, Ryu10, Kitaev09, Chiu16}, in the sense that the definition of topological invariants in terms of Bloch states and what measurable quantities they correspond to have all been thoroughly investigated~\cite{Hasan10, Qi11}. 

In fact, it has been pointed out recently that all the Dirac models in any dimension and symmetry class can be ubiquitously described by a single topological invariant that counts the number of times the Brillouin zone (BZ) torus wraps around a target sphere as induced by the Dirac Hamiltonian~\cite{vonGersdorff21_unification}, yielding a  simple way to calculate the topological invariant in momentum space. Moreover, a universal topological marker can be derived from this wrapping number, offering a generic way to calculate the topological invariant directly from lattice models using the lattice eigenstates~\cite{Chen23_universal_marker}. The advantage of this type of topological marker formalism is that it allows direct investigation into the effect of real space inhomogeneity, as has been widely demonstrated in the literature~\cite{Loring10, Prodan10, Prodan10_2, Prodan11, Bianco11, Bianco13, MondragonShem14, Marrazzo17, Cardano17, Meier18, Huang18, Huang18_2, Haller2020, Focassio21, Sykes21, Jezequel22, Wang22, Hannukainen22}. 

On the other hand, when many-body interactions are present, one would expect that none of the aforementioned methods of calculating the topological invariants would work since the Bloch state, lattice eigenstate, and the Dirac Hamiltonian are no longer valid concepts, such that one would need to resort to other methods. Focusing on one-dimensional (1D) systems, various definitions have been proposed instead. For instance, the degeneracy of the entanglement spectrum is found to be a robust topological invariant~\cite{Pollmann10, Turner11, Pollmann12}, and the spectrum either splits or crosses each other at the topological phase transitions~\cite{vanNieuwenburg18}, giving rise to the notion of symmetry protected topological (SPT) phases~\cite{Gu09, Senthil15}. In addition, it is also possible to construct the invariant from the Green's function in momentum space provided the interaction can be treated perturbatively~\cite{Gurarie11, Essin11}, whose singular behavior near topological phase transitions can be used to extract critical exponents~\cite{Chen18}.  

In this paper, we demonstrate that a method proposed for noninteracting 1D systems can be directly adopted to describe the topological phases in the presence of interactions, namely the topological marker. We demonstrate this feature, particularly for the spinless Su-Schrieffer-Heeger (SSH) model that belongs to the symmetry class BDI~\cite{Su79}, where the many-body ground state in the presence of nearest-neighbor (NN) interaction is solved utilizing exact diagonalization. For instance, the bosonic version of this model has been studied \cite{Fraxanet2022,Jin2023}.  We show that although the lattice eigenstates are no longer well-defined in the presence of interactions, the projectors to the filled and empty states that are key to the topological marker formalism~\cite{Bianco11, Bianco13} can still be implemented from the real space single-particle Green's function of the many-body ground state, allowing the topological marker to be introduced. In addition, adopting twisted boundary conditions to reduce finite-size effects in the ground state~\cite{Poilblanc1991}  does not preclude using the marker. 

The resulting marker in our formalism is a quantity that remains quantized in the noninteracting limit and changes continuously as the interaction is turned on. Although the non-integer marker seems to imply that the aforementioned picture of topological order as the number of times the BZ wraps around a target sphere is no longer valid in the presence of interactions, our marker still serves as a faithful tool to describe the topological phases in the sense that it remains finite in the topologically nontrivial phase and approaches zero in the thermodynamic limit within the trivial phase. In addition, our marker also correctly captures the quantum phase transition between the topologically nontrivial phase and a topologically trivial charge density wave phase, similar to what occurs in a two-dimensional (2D) correlated Chern insulator~\cite{Varney2010, Varney2011, Kourtis14, Shao2021}, as also confirmed by the calculation of the entanglement entropy and structure factor. Finally, we elaborate that our real space Green's function formalism can be generalized to lattice Dirac models of TIs in any dimension and symmetry class in the presence of any interactions and lay out the formalism for future applications.

\section{Model and Methods}
\label{sec:M&M}

\begin{figure}[t]
    \centering
    \includegraphics[scale=0.5]{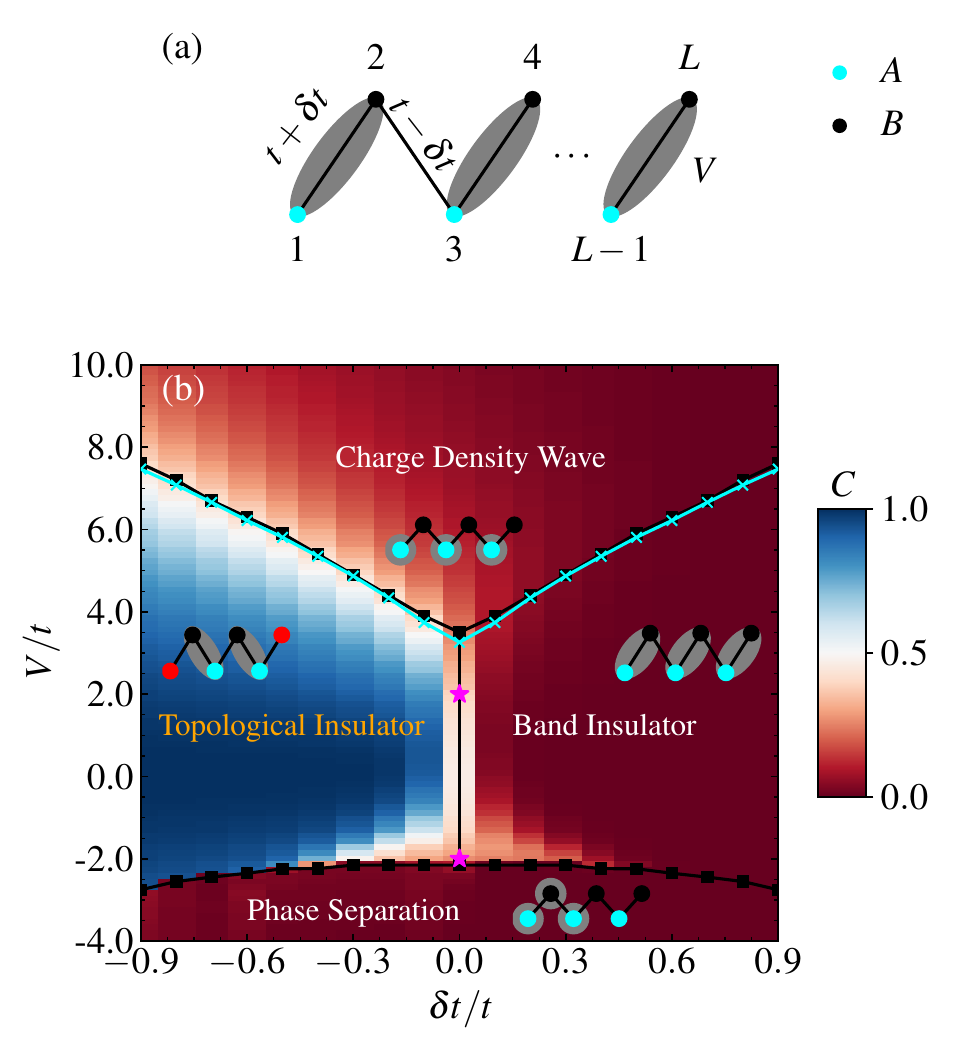}
    \caption{Schematic Hamiltonian cartoon (a) and phase diagram of the one-dimensional SSH model with NN interactions (b). The local topological marker $C \equiv C(L/4)$ is presented as a color map in the space of parameters $(\delta t, V)$, for an $L = 20$ chain where TABCs are utilized. The phase diagram indicates the existence of four phases: TI and BI, mainly governed by the dimerization $\delta t$, and CDW and PS, driven by the repulsive and attractive interactions $V$, respectively. A set of complementary quantities help in identifying the phase boundaries. In particular, the CDW transition points are identified by the linear extrapolation of the minima of $\dv{S_{vN}}{V}$ [black markers, see Fig.~\ref{fig:svn_tbc}(b)], or by the scaling of the CDW structure factor [cyan markers, see Fig.~\ref{fig:scdw_scaling} for $\delta t/t = -0.5$, as an example]. The magenta stars indicate the transition points obtained exactly for the correspondent XXZ model (via Jordan-Wigner transform) \cite{Rigol_review_2011,Rigol_2011_PRB84}, at $\delta t = 0$. For sufficiently attractive interactions, the topological marker indicates a transition for the PS regime, where the phase boundary is obtained by steep drop location of $S_{vN}$ for $V < 0$ [see Fig.~\ref{fig:svn_tbc}(a)]. The insets are schematic representations of each interacting SSH model phase, with  $A$ and $B$ sites depicted in cyan and black, respectively, as shown in (a). The gray regions indicate the unit cell. Hopping between neighboring sites within the cell takes the value  $(t - \delta t)$ while hopping between neighboring cells is $(t + \delta t)$; the NN interaction, $V$, is site-independent.}
    \label{fig:fig_1}
\end{figure}

We study the one-dimensional spinless Su-Schrieffer-Heeger  model \cite{SSH_PhysRevLett.42.1698,SSH2_PhysRevB.22.2099} with NN interactions \cite{Fradkin_PhysRevB.27.1680, Sirker_2014, Sacramento_PhysRevB.94.245123, Yahyavi_2018, Zegarra2019},
\begin{eqnarray}
    \hat {\cal H} = \sum_{\langle i,j \rangle}\left(-t + \delta t (-1)^i \right)(\hat c_{j}^{\dagger}\hat c_i^{\phantom{\dagger}} + \text{H.c.})  + V\sum_{\langle i,j \rangle}\hat n_i \hat n_j .
    \label{eq:Hamiltonian}
\end{eqnarray}
Here, $\hat c^{\dagger}_i(\hat c_i^{\phantom{\dagger}})$ denotes the fermionic creation (annihilation) operator, where $i$ and $j$ denote sites of the lattice with length $L$, with $\ave{i,j}$ restricting the sums to NN sites. $t$ represents the hopping integral, which is modulated by the dimerization $\delta t$, and $V$ is the nearest-neighbor interaction. $\hat n_i = \hat c^{\dagger}_i \hat c^{\phantom{\dagger}}_i$ is the number operator at the site $i$, with sites of a unitary cell differentiated by subscripts $A$ and $B$ as indicated in Fig.~\ref{fig:fig_1}(a).

We use the Lanczos method ~\cite{Newman_Lanczos_doi:10.2514/3.5878, Lin1993, Sandvik_Lanczos} to obtain the ground-state of the interacting SSH model [Eq.~\eqref{eq:Hamiltonian}] at half-filling. To mitigate finite-size effects, we adopt twisted-averaged boundary conditions (TABC) \cite{Poilblanc1991, Gros1992, Lin_PhysRevE.64.016702, Zawadzki2017}. 
Thus, whenever a fermion hops between the first and last sites of the system, the hopping gains a phase $\phi$:  $t_{ij} \rightarrow t_{ij}\exp{{\rm i}\phi_\ell}$,
where $\phi_\ell = [0,2\pi)$. Each $\phi_\ell$ allows a different set of $k$-points, increasing the number of available momenta in the first Brillouin zone with a given lattice and reducing finite-size effects. In this framework, the expected value of an operator $\hat{A}$ is given by~\cite{Koretsune2007}
\begin{equation}   
\langle \hat{A} \rangle = \frac{1}{N_{\phi}}\sum_{\ell = 1}^{N_{\phi}}\langle \hat{A}\rangle_{\phi_\ell}\ ,
\end{equation}
where we set $N_{\phi} = 20$ as the number of values of $\phi_\ell$.
 
To probe if the system is in a metallic or insulating state, we calculate the energy gap between the ground and the first excited states, $\Delta \equiv E_1 - E_0$. We further calculate charge-charge correlation functions since repulsive interactions enhance these and can display long-range orders for sufficient $V$. Due to the dimerization $\delta t$, it is convenient to divide the chain into two sub-lattices, $A$ (odd sites) and $B$ (even sites), as indicated in Fig.~\ref{fig:fig_1}(a). This separation is useful to define NN intra-cell $\langle \hat n_{i\in A}\hat n_{i\in B}\rangle$ and inter-cell $\langle \hat n_{i+1\in A}\hat n_{i\in B}\rangle$  correlation functions.
The charge density wave structure factor is defined as 
\begin{equation}
    S_{\text{cdw}} = \frac{1}{L}\sum_{i,j}\langle \hat n_{i}\hat n_{j}\rangle (-1)^{\abs{i - j}}\ ;
    \label{eq:S_cdw}
\end{equation}
long-range order entails that $S_{\rm cdw}$ is extensive in the system size. In turn, the von Neumann entropy,  $S_{vN}$, measures the entanglement between two parts, $A$ and $B$ of a system and is given by 
\begin{equation}
    S_{vN}(\delta t, V) = -\Tr{\rho_A \ln{\rho_A}},
    \label{eq:vN_entropy}
\end{equation}
here we have chosen $A$ to be the set of sites in sub-lattice $A$ and $B$ the set of sites in sub-lattice $B$. 
So $\rho_A$ is the partial trace of the density matrix $\rho$ over the sub-lattice $B$. 

The characterization of topological phases is not trivial for the interacting SSH model, as a solution by Fourier transform is unavailable due to the presence of correlations. We investigate the model's topology using the local topological marker, $C(r)$~\cite{Zak_Resta_2000, Chen23_universal_marker}, which provides insight into the system's topology in real space.  $C(r)$ is defined within each unit cell of the lattice, and for non-interacting systems it is given by
\begin{equation}
    C(r) = \sum_{i = \{r,A\}, \{r,B\}} \bra{i}\hat \sigma_z (\hat P\hat X\hat Q + \hat Q\hat X\hat P)\ket{i},
    \label{eq:chern_local}
\end{equation}
where $\nu$ is the sub-lattice index,  $\hat \sigma_z$ is the $z$ Pauli matrix, $\hat P \equiv \ket{\Psi_{0}}\bra{\Psi_{0}}$ is the ground-state projector, $\hat Q = \sum_{m > 0} \ket{\Psi_{m}}\bra{\Psi_{m}}$ is the projector on the excited states and $\hat X$ is the unit cell position operator. We can rewrite that equation by noticing that the matrix elements of $\hat P$ are equivalent to the single-particle Green's function $\hat G$, with components $G_{i,j} = \langle \hat c_{i}^{\phantom{\dagger}}\hat c_{j}^{\dagger}\rangle$, and $\hat Q = (\hat I - \hat G)$ by completeness. Therefore, we can rewrite $C(r)$ as
\begin{equation}
    C(r) = \sum_{i = \{r,A\}, \{r,B\}} \bra{i}\hat \sigma_z (\hat G \hat X(\hat I - \hat G) + (\hat I - \hat G)\hat X\hat G)\ket{i}.\label{chern_greens_local}
\end{equation}

Finally, we also calculate the many-body Zak phase \cite{Zak_PRL.62.2747} given by the expression
\begin{equation}
\gamma = -\Im\log \prod_{\ell = 0}^{N_{\phi}}\braket{\psi(\phi_{\ell})}{\psi(\phi_{\ell+1})},
\label{eq:ZP_manybody}
\end{equation}
where $\ket{\psi(\phi_{\ell})}$ is the ground state of the SSH model for twisted boundary conditions with a phase $\phi_\ell$. 

\section{General formalism for topological insulators in arbitrary dimension}

We anticipate that our formalism of topological marker that replaces the projectors by the real space Green's function is valid to the SSH model and generally applicable to lattice Dirac models of TIs in any dimension and symmetry class. This conjecture is made because the universal topological marker applicable to any dimension and symmetry class also takes the form of alternating projectors and position operators as in Eq. \ref{eq:chern_local}, and hence the Green's function formalism should be directly applicable \cite{Chen23_universal_marker}. Below we explicitly outline such a formalism.

Consider a $D$-dimension TI described by the Dirac Hamiltonian in momentum space $H_{0}({\bf k})={\bf d}({\bf k})\cdot{\boldsymbol\Gamma}$, where $\Gamma_{i}=(\Gamma_{0},\Gamma_{1}...\Gamma_{2n})$ are the $n$-th order Dirac matrices of dimension $2^{n}\times 2^{n}$ that satisfy the Clifford algebra  $\left\{\Gamma_{i},\Gamma_{j}\right\}=2\delta_{ij}$, and ${\bf d}({\bf k})=(d_{0},d_{1},...,d_{D})$ describes the momentum dependence of the Hamiltonian. The real space lattice Hamiltonian $\hat H_{0}$ can be straightforwardly obtained from a Fourier transform of $\hat H_{0}({\bf k})$, whose basis contains the electron operators $\hat c_{i}$ with $i=\left\{{\bf r},\nu\right\}$, where ${\bf r}$ denotes the position of the unit cell, and $\nu$ enumerates any internal degrees of freedom such as spin, orbital, sublattice, etc. We denote the interacting Hamiltonian by $\hat H_{\rm int}$ such that the full Hamiltonian is $\hat H=\hat H_{0}+\hat H_{\rm int}$. After the many-body ground state $|\psi\rangle$ is obtained numerically, we aim to calculate the universal topological marker in terms of $|\psi\rangle$. This is done by considering a universal topological operator constructed out of $\hat H_{0}$ that takes the form of two arrays of alternating projectors $\left\{\hat P,\hat Q\right\}$ sandwiched by position operators ${\hat r}_{1\sim D}=\left\{{\hat x},{\hat y},{\hat z}...\right\}$
\begin{eqnarray}
{\hat {\cal C}}=N_{D}\hat W\left[\hat Q\,{\hat r_{1}}\hat P\,{\hat r_{2}}...\,{\hat r_{D}}\hat {\cal O}+(-1)^{D+1}\hat P\,{\hat r_{1}}\hat Q\,{\hat r_{2}}...{\hat r_{D}}{\overline{\hat{\cal O}}}\right],
\nonumber \\
\label{topological_operator}
\end{eqnarray}
where $\hat W=\hat \Gamma_{D+1}\hat \Gamma_{D+2}...\hat \Gamma_{2n}$ is the product of all the unused Dirac matrices. In this expression, the last operators $\left\{\hat {\cal O},\overline{\hat{\cal O}}\right\}=\left\{\hat P,\hat Q\right\}$ if $D=$ odd, and $\left\{\hat {\cal O},\overline{\hat {\cal O}}\right\}=\left\{\hat Q,\hat P\right\}$ if $D=$ even owing to the alternating ordering of the projectors $\hat Q$ and $\hat P$. The normalization factor is $N_{D}={\rm i}^{D}2^{2D-n}\pi^{D}/c\,V_{D}$, with $V_{D}=\left\{V_{1},V_{2},V_{3}...\right\}=\left\{2\pi,4\pi,2\pi^{2}...\right\}$ the volume of a $D$-sphere, and the prefactor $c={\rm Tr}\left[\hat \Gamma_{0}\hat \Gamma_{1}...\hat \Gamma_{2n}\right]/2^{n}=\left\{1,-1,{\rm i},-{\rm i}\right\}$ depends on the representation of $\Gamma$-matrices for the system at hand. 

In the noninteracting limit $\hat H_{\rm int}=0$, the projectors $\hat P$ and $\hat Q$ in Eq.~(\ref{topological_operator}) are evaluated by the summations of projectors to the filled and empty lattice eigenstates, respectively \cite{Bianco11, Chen23_universal_marker}. Encouraged by our results in the interacting SSH model, we propose that for interacting systems $\hat H_{\rm int}\neq 0$, the projectors $\hat P\rightarrow \hat G$ and $\hat Q\rightarrow \hat I-\hat G$ may be replaced by the single-particle Green's function $G_{i,j}=\langle \hat c_{i}\hat c_{j}^{\dag}\rangle$, where the expectation value $\langle \ldots\rangle$ may contain average over twisting angles if the twisted boundary condition is adopted. The topological marker at unit cell ${\bf r}$ is then calculated by
\begin{equation}
C({\bf r})=\sum_{\nu}\langle{\bf r},\nu|{\hat {\cal C}}|{\bf r},\nu\rangle,
\end{equation}
similar to that in Eq.~(\ref{eq:chern_local}). This topological marker should readily apply to any lattice Dirac models of TIs with many-body interactions, which is left for future investigations. Finally, we remark that another intriguing issue is whether this Green's function formalism is also applicable to interacting TSCs, whose noninteracting limit can also be described by Eq.~(\ref{topological_operator}). However, because the basis of the TSCs involves both electron creation and annihilation operators, the Green's function will contain the anomalous components like $\langle \hat c_{i}\hat c_{j}\rangle$ and $\langle \hat c_{i}^{\dag}\hat c_{j}^{\dag}\rangle$, and it remains unclear to us at present how they enter Eq.~(\ref{topological_operator}) appropriately. This issue awaits to be further explored. 

\begin{figure}[t]
\includegraphics[scale=0.5]{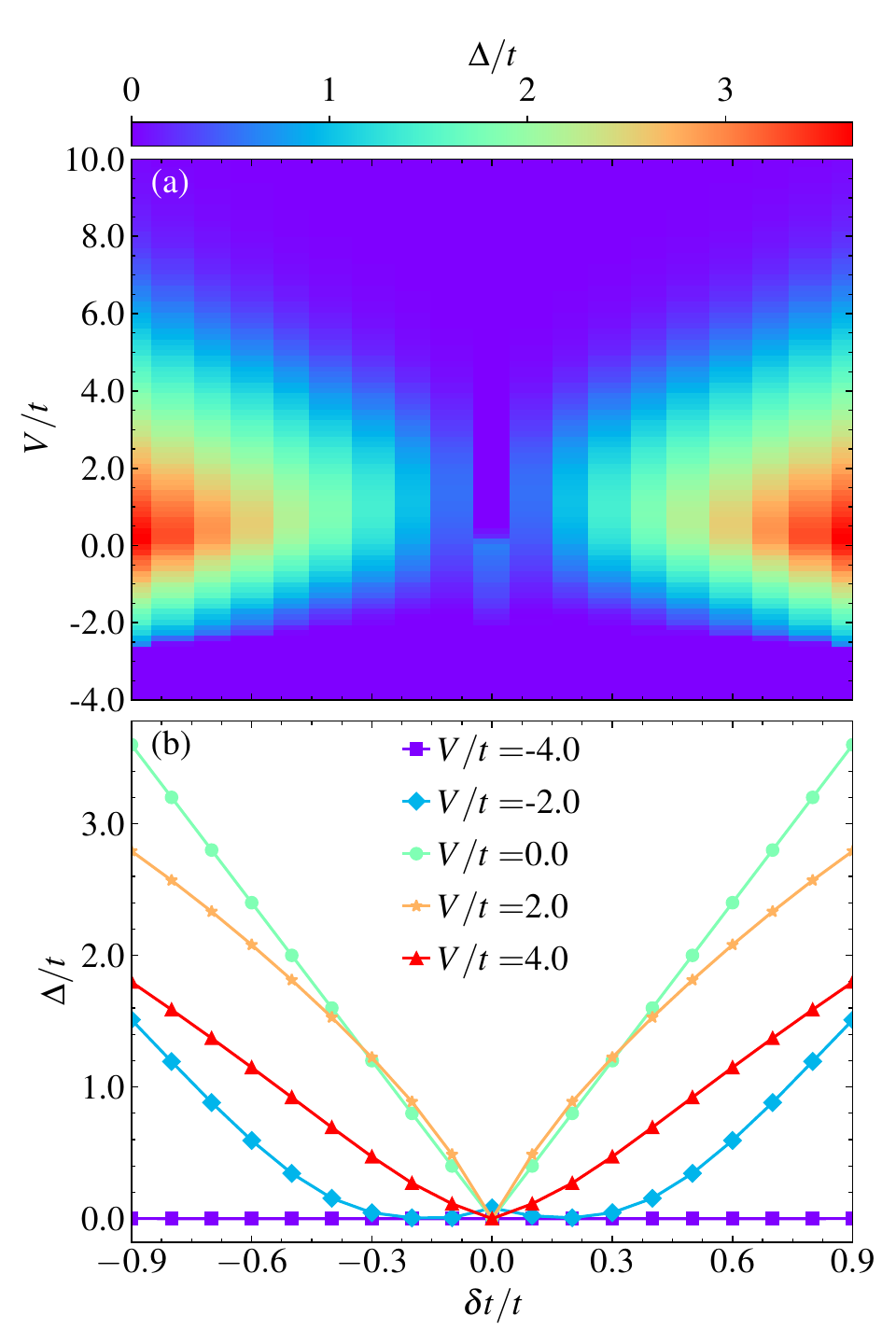}
\caption{\label{fig:gap_vs_deltat} Energy gap in the $(\delta t, V)$ parameters space, as a color map in (a) and corresponding line cuts at particular interaction strengths in (b). Here, data is extracted at a fixed lattice size $L=20$, and averaged over twisted boundary conditions. Topological and band-insulating regions exhibit a well-marked gap to excitations, unlike the phase-separated and Mott regions (see text).}
\end{figure}

\section{Results}\label{int_limit}
Having established the main quantities, we now systematically characterize the different phases of \eqref{eq:Hamiltonian}, as described in Fig.~\ref{fig:fig_1}(b). Other than a phase-separated regime occurring at sufficiently negative interactions $V$ on a broad range of dimerization values $\delta t$, three extra insulating phases emerge, topological (band) insulator at moderate interaction strengths for $\delta t < 0$ ($\delta t > 0$), and a (topologically trivial) Mott insulator featuring a charge density wave when $V$ is large. As will become clear in what follows, such a transition is characterized by a $Z_2$ symmetry breaking, thus being governed by the 2d-Ising universality class in the interacting SSH chain. The different quantitities introduced in Sec.~\ref{sec:M&M} will be individually used to establish the corresponding phases.

\subsection{Energy gap} \label{sub:gap}

We start by classifying the energy gap, i.e., the dependence of the excitation energy between the ground-state and the first excited one in the dimerization $\delta t/t$ for different values of $V/t$, as shown in Fig.~\ref{fig:gap_vs_deltat}(a). Comparing it with the phase diagram in Fig.~\ref{fig:fig_1}(b), one notices that both the topological and band insulating regions display a robust gap. This extends to the interacting realm of the known finite gaps obtained in the $V=0$ case (see Appendix~\ref{app:AppendixA}). Furthermore, the direct TI$\leftrightarrow$BI transition is then characterized by the gap closure for interactions $0 \lesssim V \lesssim 3t$ at $\delta t = 0$.

\begin{figure}[t]
\includegraphics[scale=0.5]{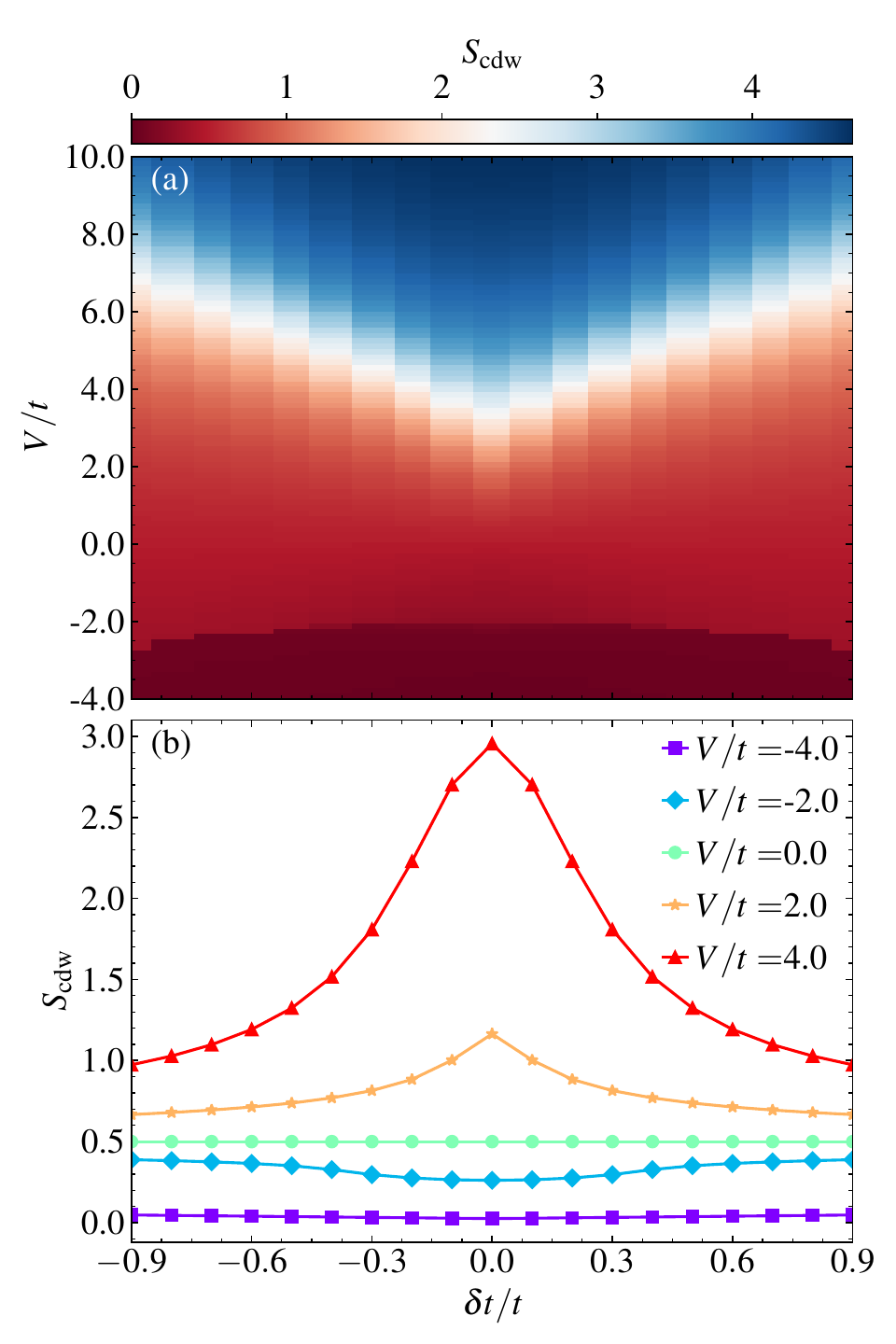}
\caption{\label{fig:scdw_vs_deltat} CDW structure factor $S_{\rm cdw}$ in the $(\delta t, V)$ parameters space, shown as a colormap in (a) and line cuts along representative interaction values in (b). As in Fig.~\ref{fig:gap_vs_deltat}, data is extracted for an $L=20$ lattice at half-filling and averaged over twisted averaged boundary conditions. The MI region is characterized by a robust $S_{\rm cdw}$ value.}
\end{figure}

\begin{figure}[t]
\includegraphics[scale=0.5]{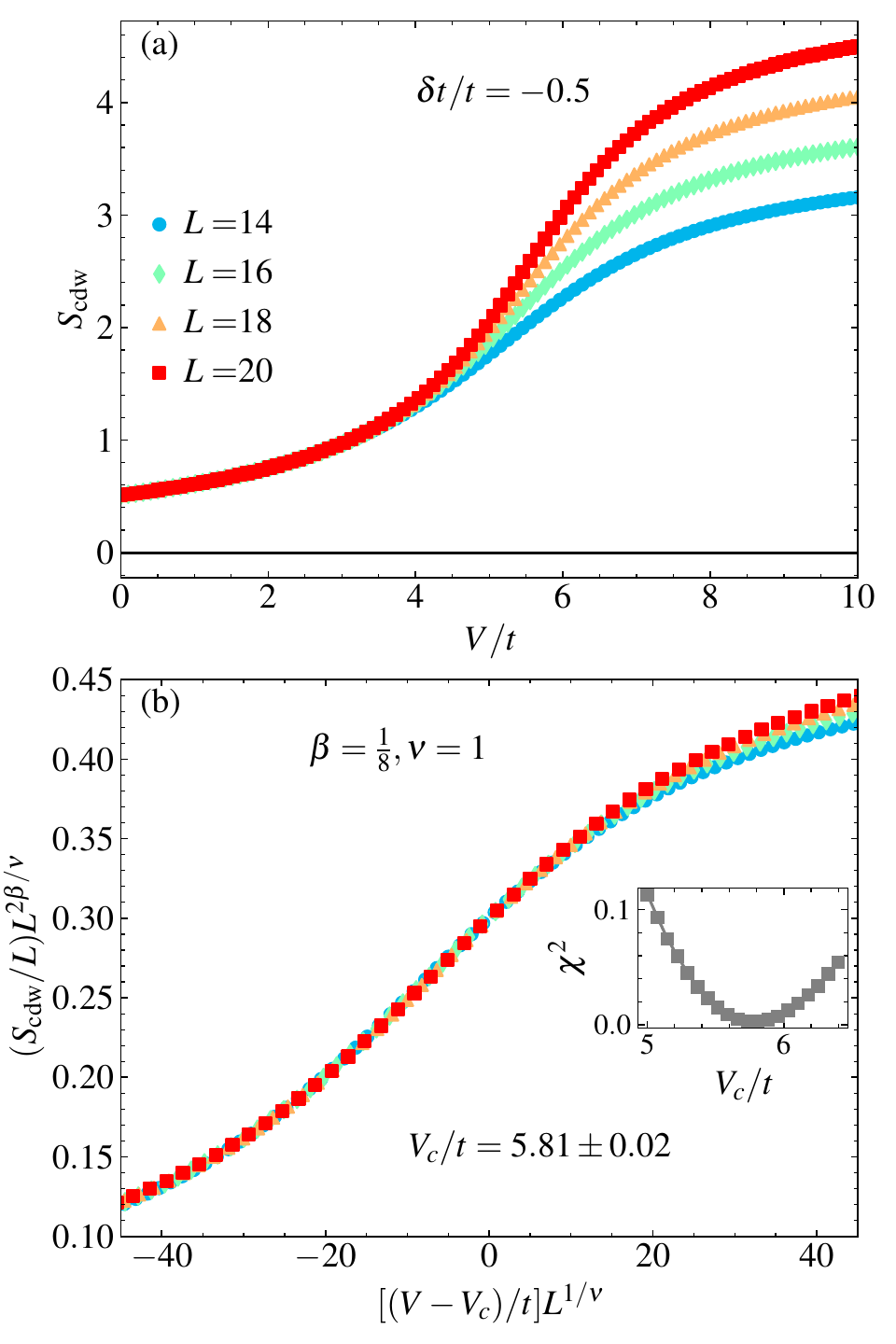}
\caption{\label{fig:scdw_scaling} (a) The $V$-dependence of the CDW structure factor with $\delta t = -0.5t$ for various system sizes, and (b), the scaled $S_{\rm cdw}$ according to the scaling ansatz [Eq.~\eqref{eq:scaling_ansatz}]. Here, the critical interaction $V_c^{\rm cdw} = 5.81 \pm0.02$ triggers the CDW phase for this value of $\delta t$, and is the one that minimizes the $\chi^2$ of the collapse (inset); see text.
}
\end{figure}

Conversely, the phase-separated and Mott insulator regions are characterized by a vanishing of the energy gap only in the thermodynamic limit. While in the former, an extensive number of states exhibiting various configurations of phase separation display similar energy in the low-lying spectrum (thus making $\Delta \to 0$), in the latter, there is a doublet of charge-density-wave states (even and odd under inversion symmetry) which become degenerate in approaching the atomic limit ($t/V\to 0$).

These two considerations establish that while the direct transition from the topological insulator to the trivial band insulator is a first-order phase transition characterized by an energy level crossing and thus gap closure, the TI$\leftrightarrow$MI is a typical second-order phase transition, whose universality becomes clear when analyzing the emergent order parameter in what follows. A precise characterization of the gaps when approaching $L\to\infty$ is given in Appendix~\ref{app:AppendixB}.

\subsection{Charge correlations and charge density wave}

Given that the interactions $V$ are the dominant energy scale, one expects a CDW state in the bipartite SSH chain to occur. The precise value of the interaction strength that triggers this Mott insulating state with a finite local order parameter depends on the magnitude of the dimerization $\delta t$. Figure~\ref{fig:scdw_vs_deltat}(a)  maps out the CDW structure factor $S_{\rm cdw}$ on the $(\delta t, V)$ plane. As predicted, once $V \gg t, \delta t$ a large $S_{\rm cdw}$ identifies a favored CDW regime, whose critical value of the interactions $V_c$ to necessary to trigger it grows roughly linearly with  $\abs{\delta t/t}$. Furthermore, such a transition occurs irrespective of whether the parent regime is topologically trivial or not. In particular, the onset of the CDW phase at vanishing dimerization occurs at values of the interactions $V/t \gtrsim 3$.

\begin{figure}[t]
\includegraphics[scale=0.5]{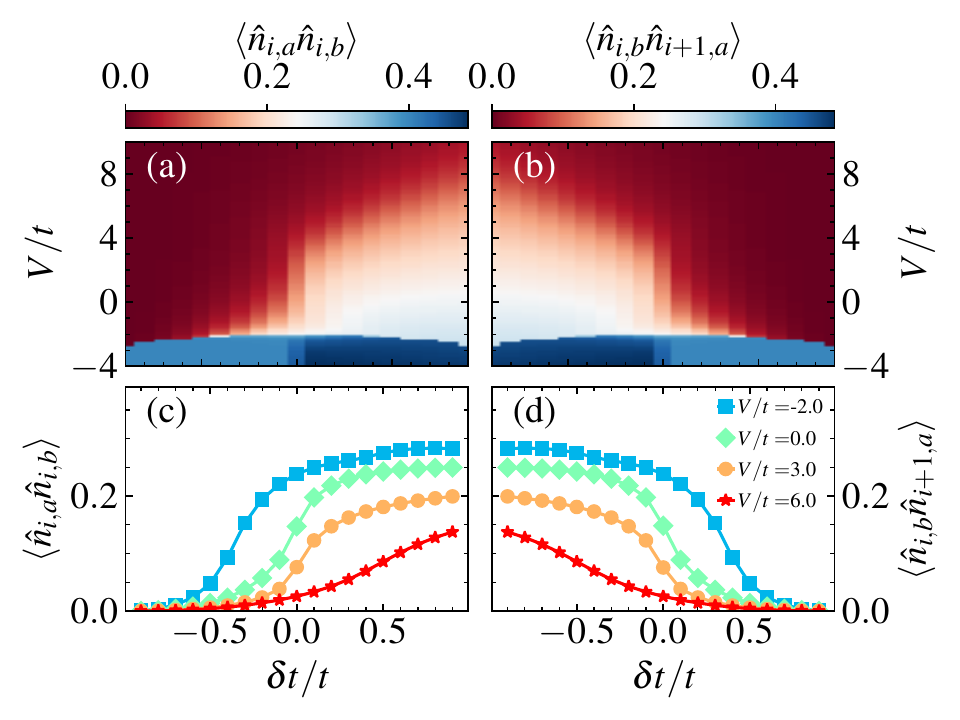}
\caption{\label{fig:nn_functions} Color map of the intra- and intercell nearest-neighbor correlation functions, (a) and (b), respectively, in the space of parameters $(\delta t, V)$. Panels (c) and (d) depict the same for selected interaction strength values $V$. As in previous figures, data refers to a $L=20$ lattice averaged over twisted boundary conditions.}
\end{figure}

While suggestive, only a precise scaling of $S_{\rm cdw}$ can confirm the manifestation of a phase transition to the ordered regime. We start by noticing that a CDW state is tied to a spontaneous symmetry breaking of the sublattice symmetry in the thermodynamic limit, thus being classified as a continuous phase transition identified by a $Z_2$-symmetry breaking. If defining the order parameter $m_{\rm cdw} \equiv |\langle \hat n_A - \hat n _B\rangle|$, we notice that the structure factor [Eq.~\ref{eq:S_cdw}] is proportional to $m_{\rm cdw}^2 L$. Assuming thus that, given the symmetry-breaking form, such phase transition pertains to the $(1 + 1)-d$ Ising universality class, $S_{\rm cdw}/L$ should obey the scaling ansatz,
\begin{equation}
    (S_{\rm cdw}/L) = L^{-2\beta/\nu} g[((V - V_{c})/t)L^{1/\nu}],
    \label{eq:scaling_ansatz}
\end{equation}
where the exponents $\beta = 1/8$ and $\nu = 1$ are the characteristic ones for this universality class.

\begin{figure*}[t]
\includegraphics[scale=0.5]{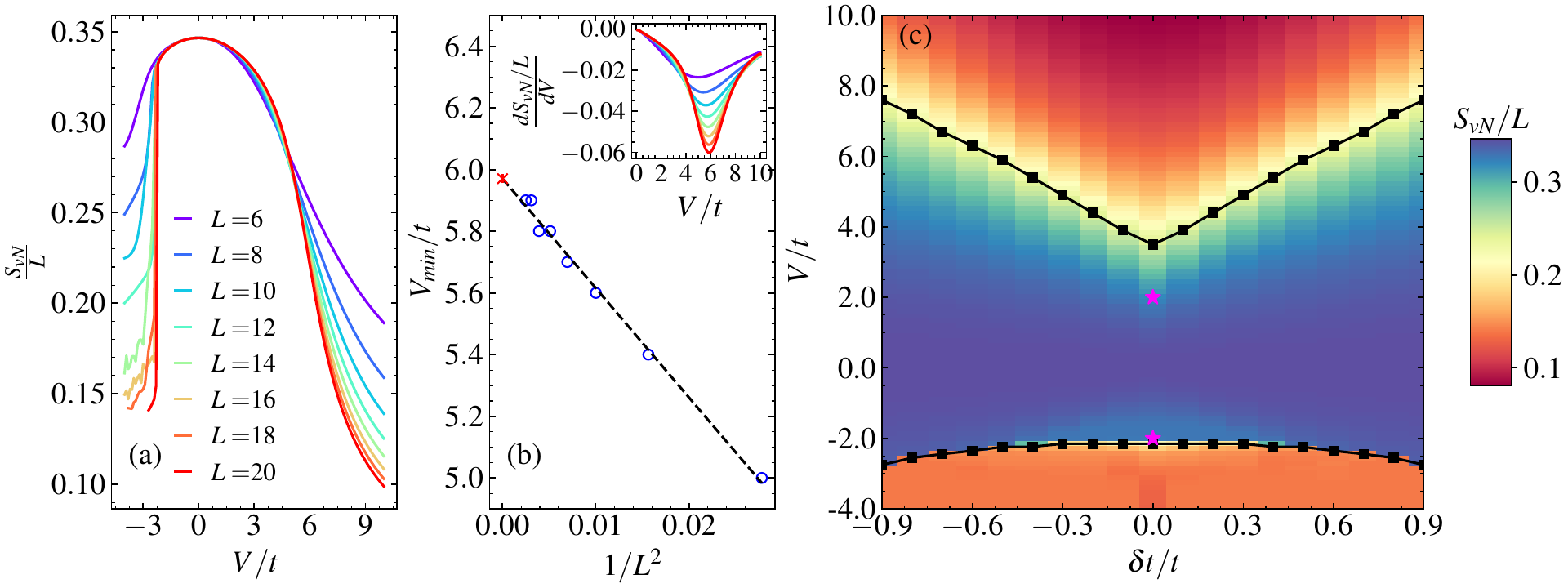}
\caption{\label{fig:svn_tbc} 
(a) $S_{vN}$ as a function of $V/t$ for $\delta t = -0.5$ for chains with $L = 6, 8, 10, 12, 14, 16, 18, 20$, and (b), the corresponding extrapolation of the minima position $V_{\rm min}/t$ of $\dv{S_{vN}}{V}$ with $1/L^2$; the inset shows $\dv{S_{vN}}{V}$ vs.~$V/t$ at the same parameters as in (a). Panel (c) displays a colormap of the von Neuman entropy in the ($\delta t, V$) space of parameters at a fixed lattice size $L=20$. The magenta stars in (c) indicate the transition points obtained exactly for the correspondent XXZ model (via Jordan-Wigner transform) \cite{Rigol_review_2011,Rigol_2011_PRB84}, at $\delta t = 0$. }
\end{figure*}

Following this rationale, Fig.~\ref{fig:scdw_scaling} shows both the extensive behavior of $S_{\rm cdw}$ within the CDW phase and the scaling form satisfying the ansatz at $\delta t/t = -0.5$. The critical interaction $V_c$ is estimated by the minimum of a $\chi^{2}$-test, quantifying the scaling collapse. It is written as $\chi^{2}_{V_{c}, \beta, \nu} = \sum_{j}(x_{j} - m_{j})^2/m_{j}$, where $x_{j}$ are the values of $(S_{\rm cdw}/L)/L^{2\beta/\nu}$ for each $(V - V_{c})L^{1/\nu}$, and $m_{j}$ is the expected power law fit for the curve. A compilation of this analysis for different values of $\delta t$ gives the cyan empty markers in Fig.~\ref{fig:fig_1} showing the critical interactions $V_c$.

While this scaling analysis unequivocally establishes the emergence of charge ordering at large interaction strengths, a direct inspection of the nearest-neighbor (intra and inter-unit cell) correlations gives extra insight into the different phases, whether charge-ordered or not. Figures~\ref{fig:nn_functions} (a,c) and \ref{fig:nn_functions}(b,d) report the dependence of both intra-, $\langle\hat{n}_{i,A},\hat{n}_{i,B}\rangle$, and inter-cell, $\langle\hat{n}_{i,B}\hat{n}_{i+1,A}\rangle$, NN correlation functions on the space of parameters $(\delta t, V)$. 

The non-interacting regime ($V=0$) reproduces known behavior, namely, the existence of topological dimers (inter-cell) for $\delta t < 0$ and trivial dimers (intra-cell) for $\delta t > 0$. Introducing a finite repulsive interaction lowers the curves of both correlation functions, indicating that the presence of the interaction term $V$ competes with the dimerization term $\delta t$, eventually breaking the dimer regime. This result is consistent with the subsequent formation of a CDW phase once $V$ is sufficiently large. In turn, large attractive interactions result in intra- and inter-cell NN correlations, which are both substantially large [see blue regions in Figures~\ref{fig:nn_functions} (a,b)], a signature of the charge accumulation characteristic of phase-separated regimes. 

The findings suggest that the interplay between dimerization $\delta t$ and interactions $V$ can lead to the emergence of new phases, in addition to the well-known TI and BI phases. Notably, a CDW phase in the repulsive regime, particularly for strong interaction strengths, and a phase-separated one for strong, attractive interactions. The competition between these two `knobs' directly explains the roughly linear dependence of $V_c$ on $\abs{\delta t}$ [see Fig.~\ref{fig:fig_1}(b)].

\subsection{von Neumann entanglement entropy \label{SvN_section}}

We next examine the von Neumann entropy, recalling that the partitions $A$ and $B$ used to calculate $S_{vN}$ coincide with the sublattices of the SSH chain. A quantum phase transition can either be identified by a singularity in $S_{vN}$ or a minimum of the derivative of $S_{vN}$ with respect to a control parameter~\cite{Lin2004, Sirker_2014}. Figure \ref{fig:svn_tbc}(a) shows $S_{vN}/L$ for $\delta t/t=-0.5$ as a function of $V/t$ for different system sizes. In the repulsive regime, the curves of the entanglement entropy cross at a single value of the interaction strength [Fig.~\ref{fig:svn_tbc}(a)], which can also be obtained by the locus of the $\dv{S_{vN}}{V}$ singularity, when extrapolating it to the thermodynamic limit [Fig.~\ref{fig:svn_tbc}(b)]. The obtained value of $V_{c}^{vN}/t = 5.97 \pm 0.02$ is sufficiently close to the one obtained by scaling the CDW structure factor to rule out the possibility of any intermediate phase. Larger lattice sizes can potentially bring an even closer agreement.

Next, we present the results for $S_{vN}$ in Fig.~\ref{fig:svn_tbc}(c) as a color plot in the $(\delta t, V)$ parameter space. This analysis was performed for an $L = 20$ chain considering TABCs. For $V > 0$, the solid line that varies with $\delta t$ indicates the extrapolated transition points to the topologically trivial Mott insulator, obtained as above. For $V < 0$, the solid lines indicate the location of the systematic drop in ${S_{vN}}$, which, according to the previous analysis of the correlations, denotes the onset of the phase-separated regime. Since the ground state of the region exhibiting PS is a highly non-entangled object, it makes its identification increasingly sharp, with minimal finite-size effects.

\subsection{Zak's phase and topological marker}

\begin{figure}[t]
\includegraphics[scale=0.5]{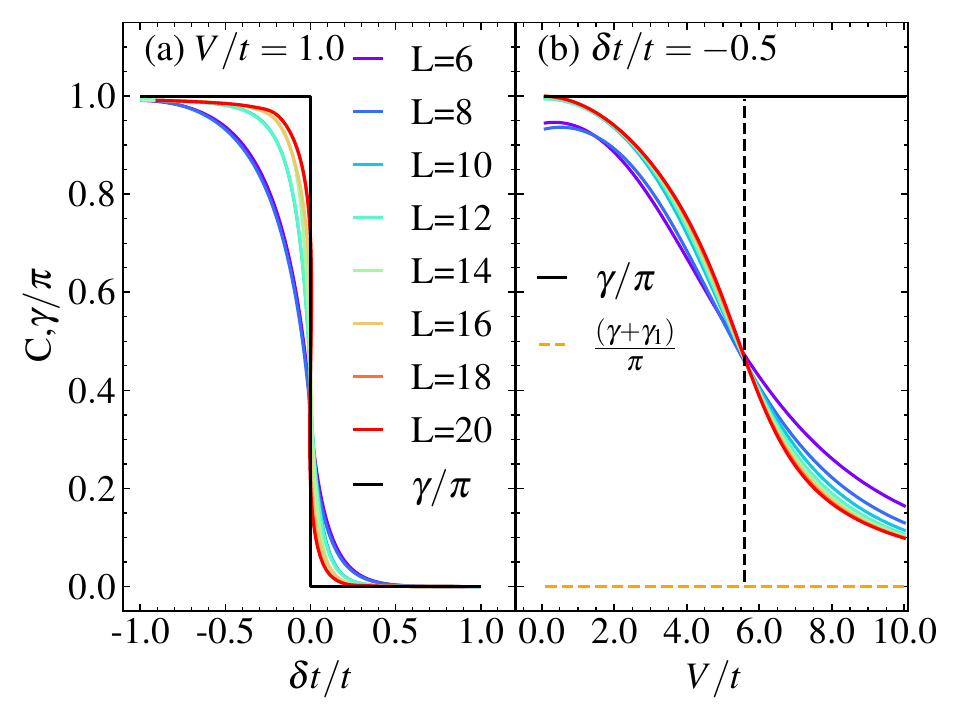}
\caption{(a) Zak's phase $\gamma$ and topological marker $C$ as a function of $\delta t$ for $V=t$ for several lattice sizes. (b) Using the same lattice sizes, the topological marker as a function of $V$ for $\delta t /t= -0.5$. The crossing point is marked by the vertical dashed line, agreeing with the location of the TI$\leftrightarrow$CDW transition as verified by the scaling of $S_{\rm cdw}$ and of the $S_{vN}$ for this dimerization value. Here the topological invariant (Zak phase) does not change but the total value when combining $\gamma$ from both the ground and first excited states vanishes -- see discussion in the text.
}
\label{fig:chern_decay_V}
\end{figure}

Finally, the topological properties can be directly obtained by the computation of the Zak phase [Eq.~\eqref{eq:ZP_manybody}], as previously done in other studies of many-body systems~\cite{Sacramento_PhysRevB.94.245123}. Nonetheless, we argue that the topological marker [Eq.~\eqref{eq:chern_local}] can extract the same information about topology on a fraction of the computational cost. Unlike Zak's phase, where one must compute the many-body ground state on a sufficiently discretized set of twisted boundary conditions, it is sufficient to use a single boundary condition for the topological marker. To exemplify this, we compute both $\gamma$ and $C(r)$ for numerous $(\delta t, V)$ values in Fig.~\ref{fig:chern_decay_V}. Here, we chose the value of $C(L/4)$ because it exhibits a quantized response in the non-interacting limit ($V = 0$)~\cite{Chen23_universal_marker}. Figure~\ref{fig:chern_decay_V}(a) indicates that already in the weakly interacting regime ($V=t$) the $C(L/4)$ steadily converges to the actual Zak phase when approaching the thermodynamic limit, but the crossing at $\delta t = 0$ is sufficient to pinpoint the TI$\leftrightarrow$BI transition. 

Now, fixing the dimerization at $\delta t = -0.5$, we notice the same crossing for $C(L/4)$ when considering different system sizes in the repulsive regime coincides with the critical interaction that triggers the CDW Mott insulating phase. As a result, reasonably small system sizes with a single boundary condition can accurately locate the topological transition when using such a marker. It is important to emphasize that if using the Zak phase in this case, we do not see a transition from $\gamma/\pi=1$ to 0 as the local topological marker seems to suggest. We recall that the spontaneous symmetry breaking that gives rise to a CDW state (breaking inversion symmetry) is only obtained in the thermodynamic limit. As a result, the excitation gap in a finite system size is always finite (see Appendix~\ref{app:AppendixB}), converging to zero at $V=V_c$ if $L\to\infty$. Such resilience to the change of the topological invariant under spontaneous symmetry-breaking perturbations has been previously reported~\cite{Raj2021}, and similarly, the change of the topological invariant takes place only when the corresponding excitation gap closes. As a result, in our finite-sized calculations, the Zak phase remains unaltered when entering the ordered regime.

Nonetheless, as established in Sec.~\ref{sub:gap}, the ground state is two-fold degenerate in the thermodynamic limit within the ordered regime. In finite lattices, $\Delta>0$, but if instead computing the combined Zak phase $(\gamma + \gamma_1)/\pi$, i.e., the sum of the topological invariant of both the ground and first excited states, one thus sees that this total topological invariant is zero [see Fig.~\ref{fig:chern_decay_V}(b)], a preliminary indication of what to expect in the thermodynamic limit.

\section{Conclusions}

We characterize the phase diagram of the interacting SSH chain model and further elaborate on the feasibility of a topological marker in identifying its topological phases. The topological marker is generalized from the noninteracting one by reinterpreting the projectors $\hat P$ and $\hat Q$ in terms of the real space single-particle Green's function of the many-body ground state, and (if needed) it can be combined with the twisted boundary condition to improve accuracy. Our results show that such a marker is not quantized in the presence of the nearest-neighbor interactions but nevertheless always remains finite in the topologically nontrivial phase while systematically decreasing with system size in the topologically trivial one. As a result, it can be used a useful proxy to describe the topological properties of the model at a fraction of the effort of Zak's phase. 

In addition, we characterized the emergence of the CDW phase for sufficiently strong interactions identifying it as a second-order phase transition within the 2$d$-Ising universality class. The topological marker also captures such a transition, departing from the TI, as confirmed by the singularity in the derivative of the entanglement entropy, suggesting that the marker is feasible even in fermionic systems that exhibit topological and Landau orders. These encouraging results lead us to conjecture that replacing the projectors with the real space Green's function should be generically applicable to the topological markers of TIs belonging to any dimension and symmetry class. Moreover, our marker also allows the interplay between strong correlations and other complications in real space, such as disorder and grain boundaries, to be investigated. In addition, it is intriguing to ask whether the marker can capture multiple transitions between different integer values of topological invariants driven by interactions, not just from $1$ to $0$ as demonstrated in the present work. These intriguing and realistic issues await to be further explored.

\begin{acknowledgments}
The authors are grateful to the Brazilian Agencies Conselho Nacional de Desenvolvimento Cient\'{ı}fico e Tecnol\'{o}gico (CNPq), Coordena\c{c}\~{a}o de Aperfei\c{c}oamento de Pessoal de Ensino Superior (CAPES), Funda\c{c}\~{a}o Carlos Chagas de Apoio \`{a} Pesquisa do Estado do Rio de Janeiro (FAPERJ), and Instituto Nacional de Ci\^{e}ncia e Tecnologia de Informa\c{c}\~ao Qu\^{a}ntica (INCT-IQ) for funding this project. W.C.~acknowledges the financial support from CNPq Grant No. 301734/2022-4. Financial support from Funda\c{c}\~{a}o Carlos Chagas Filho de Amparo \`a Pesquisa do Estado do Rio de Janeiro grant numbers E-26/200.959/2022  and E-26/210.100/2023 (TP); and  CNPq grant numbers, 403130/2021-2  and 308335/2019-8 (TP) are gratefully acknowledged. R.M.~acknowledges support from NSFC Grants No.~NSAF-U2230402, 12050410263, 11974039, and No.~12222401.
\end{acknowledgments}

\appendix

\section{The non-interacting case}\label{app:AppendixA}

\begin{figure}[h]
\includegraphics[scale=0.5]{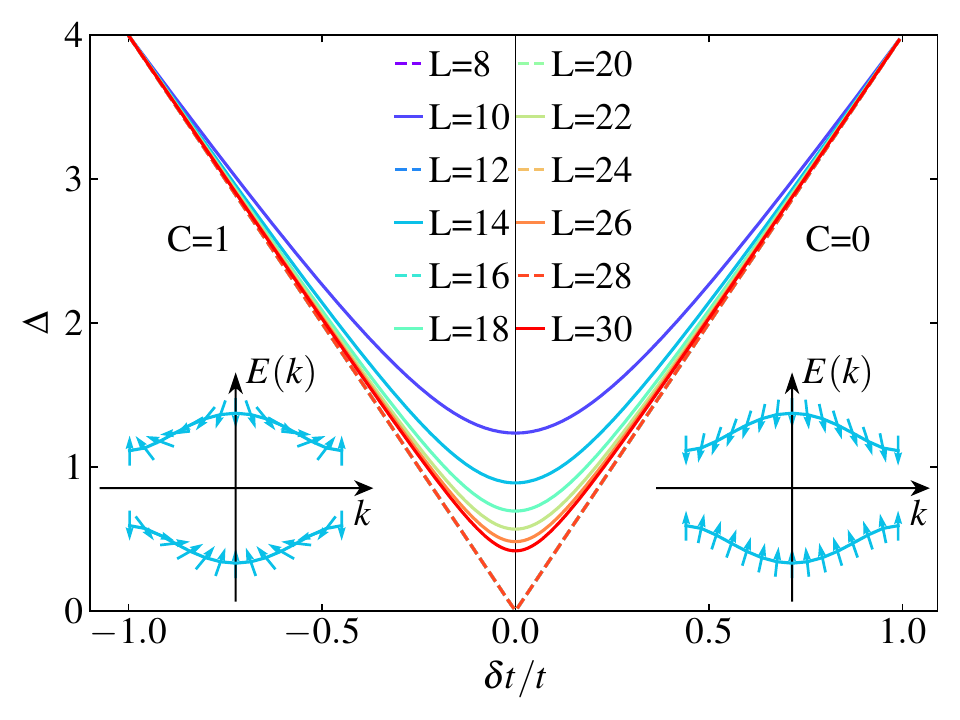}
\caption{\label{g.s.properties} Energy gap $\Delta$  as a function of the dimerization $\delta t$ for the non-interacting SSH model for chains with $L=8$  to $L=30$ sites. As expected, the gap approaches zero at $\delta t=0$ as the system's size increases. For the TI phase, the winding number completes a loop and is non-zero, as indicated in the left bottom inset. For the BI phase, the winding does not complete a loop, as shown in the right bottom inset, and therefore is zero.}
\end{figure}

In the absence of interactions ($V = 0$), it is widely recognized that two distinct phases emerge in the SSH model: the topological insulator (TI) and the band insulator (BI) \cite{Asbóth2016}. The transition between these phases is characterized by the energy level crossing between the two lowest energy states, leading to the closing of the energy gap. Additionally, the Zak phase of the lower energy band is $\gamma\color{blue}{/\pi}\color{black} = 1$ in the TI phase and $\gamma\color{blue}{/\pi}\color{black} = 0$ in the BI phase. Figure \ref{g.s.properties} displays the dependence of the energy gap $\Delta$ on the dimerization parameter $\delta t$ in the SSH model for different system sizes $L$ with periodic boundary conditions.
The gap goes to zero at $\delta t=0$ in the thermodynamic limit (the extrapolation is not shown).

In the non-interacting regime, the Hamiltonian can be written in momentum space as
\begin{equation}
    \hat {\cal H} = Q_{k}\hat c_{Ak}^{\dagger}\hat c_{Bk}^{\phantom{\dagger}} + Q_{k}^{*}\hat c_{Bk}^{\dagger}\hat c_{Ak}^{\phantom{\dagger}},
\end{equation}
where $Q_{k} = (t - \delta t) + (t - \delta t)e^{-{\rm i}k}$, and $A$ and $B$ are the different sublattices represented in Fig.~\ref{fig:fig_1}(a). 

One can define $\varphi_{k}$~\cite{Zegarra2019} by
\begin{equation}
    \varphi_{k} = -\arg(Q_{k}).
\end{equation}
for each $k \in [-\pi, \pi]$ on the first Brillouin zone. The sum $\sum_{k = -\pi}^{\pi} \varphi_{k}$ gives the winding number $\varphi$. The inset of Fig.~\ref{g.s.properties} indicates that in the TI phase, the set of $\varphi_{k}$ completes a loop, and the correspondent $\varphi$ is non-zero for $\delta t < 0$. In contrast, in the BI phase, the $\varphi_{k}$ set does not complete any loop, and $\varphi$ is zero for $\delta t > 0$.

\begin{figure}[t]
    \centering
    \includegraphics[scale=0.5]{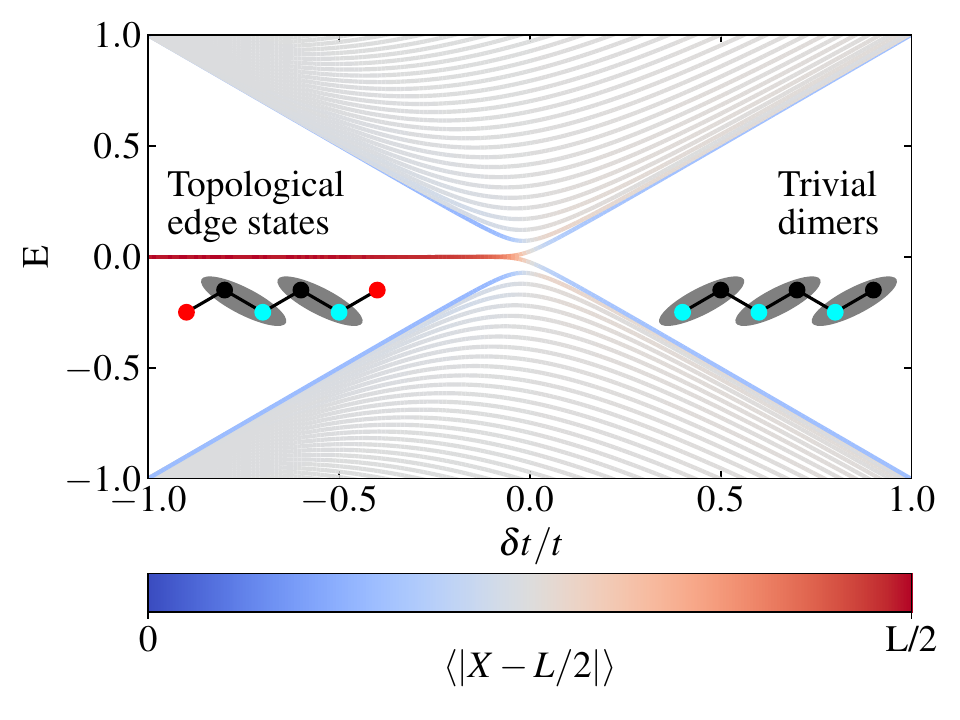}
    \caption{Energy as a  function of $\delta t / t $ for the non-interacting case for OBC. The color maps the electronic distribution of each correspondent eigenstate. It is red for states where the charge is concentrated at the edges. As the system enters the topological phase, edge states emerge in the Fermi level for $\delta t <0$. } 
    \label{en_disp_ni}
\end{figure}

\begin{figure*}[t]
\includegraphics[width=0.9\textwidth]{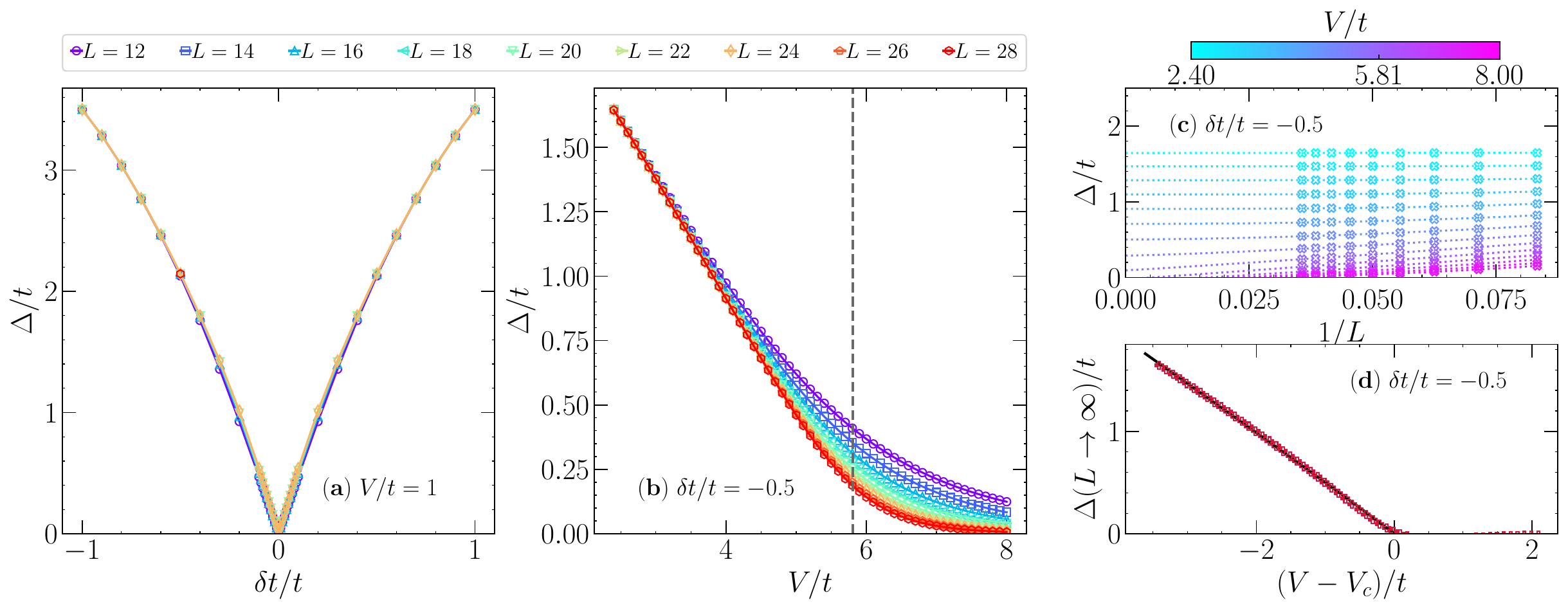}
\caption{\label{fig:gap_scaling} (a) The $\delta t$-dependence of the energy gap $\Delta$ at fixed interaction strengths $V/t=1$, indicating a typical first-order phase transition, i.e., energy-level crossing even at finite-sizes $L$. (b) The $V$-dependence of the gap at fixed dimerization $\delta t/0.5$; here, gaps are finite and only turn zero at $V=V_c$ (vertical dashed line) when approaching the thermodynamic limit, panel (c). Finally, the gap in the thermodynamic limit is a power law near the critical point, panel (d); the continuous line fits this functional form, see text.
}
\end{figure*}

The topological insulating phase at $\delta t <0$ displays a bulk-boundary correspondence that is characterized by a non-trivial topological number in the bulk and surface states at the Fermi level on the edges, as seen in  Fig.~\ref{en_disp_ni} for a $L=126$ chain with open boundary conditions. The eigenvalues of $\mathcal{H}$ are shown as a function of $\delta t/t$. For $\delta t/t <0$ two states are seen at the Fermi level. The heat map illustrates the extent to which the electronic distribution is concentrated at the edges of the chain in the TI phase. The bulk-boundary correspondence is an essential aspect of the topological properties of the SSH model, and its observation is key to the description of the non-interacting model.  

\section{Extrapolating the energy gap} \label{app:AppendixB}

In the main text, we have argued that the TI$\leftrightarrow$BI is a first-order phase transition even within the interacting regime, whereas the TI$\leftrightarrow$MI-CDW one is of second-order type. Figure~\ref{fig:gap_scaling} has the goal of establishing this quantitatively. To start, by fixing $V/t=1$, Fig.~\ref{fig:gap_scaling}(a) displays that even within finite-system sizes, the gap $\Delta\to0$ when $\delta t\to 0$, coinciding with the location of the topological transition. As a result, an unequivocal first-order phase transition emerges.

On the other hand, if fixing $\delta t/t = -0.5$, the scaling analysis of the order parameter in the main text [Fig.~\ref{fig:scdw_scaling}] argued in favor of a 2$d$-Ising-like second-order phase transition. Figure~\ref{fig:gap_scaling}(b) shows the gap dependence on the interaction strength $V$ for that fixed dimerization. While there is a clear indication of the gap reduction with increasing lattice size $L$ once it approaches the critical interaction $V_c$, these gaps remain finite. Only thus reaching the thermodynamic limit [see Fig.~\ref{fig:gap_scaling}(c)] thus one sees $\Delta(L\to\infty)\to0$ at $V=V_c$. In this limit, a typical power-law gap dependence on the distance to the critical point, $\Delta \sim |V-V_c|^{z\nu}$, ensues where $z$ and $\nu$ are the dynamic and correlation length critical exponents, respectively [Fig.~\ref{fig:gap_scaling}(d)].

\bibliography{refs}

\end{document}